\documentclass[11pt,preprint]{aastex}

\def\spose#1{\hbox to 0pt{#1\hss}}
\def\simlt{\mathrel{\spose{\lower 3pt\hbox{$\mathchar"218$}}
     \raise 2.0pt\hbox{$\mathchar"13C$}}}
\def\simgt{\mathrel{\spose{\lower 3pt\hbox{$\mathchar"218$}}
     \raise 2.0pt\hbox{$\mathchar"13E$}}}

\slugcomment{Accepted for publication in ApJL}

\shorttitle{Stellar Masses of LCBGs at $z=0.4-1.2$}
\shortauthors{Guzm\'an et al.}

\begin{document}


\title{Stellar Masses of Luminous Compact Blue Galaxies at 
Redshifts $z=0.4-1.2$}

\author{R. Guzm\'an\footnote{Department of Astronomy, University of Florida, 
       Gainesville, FL 326011, USA}, 
       G.\"Ostlin\footnote{Stockholm Observatory, SE - 106 91 Stockholm, Sweden},
       D. Kunth\footnote{Institute d'Astrophysique de Paris, 
       98 bis, Boulevard Arago, F-75014 Paris, France},
       M.A. Bershady\footnote{Department of Astronomy, University of Wisconsin, Madison, 
       475 North Charter Street, Madison, WI 53706},
       D.C. Koo\footnote{University of California Observatories/Lick Observatory, 
       Department of Astronomy and Astrophysics, University of California, 
       Santa Cruz, CA 95064},
       and 
       M.A. Pahre\footnote{Harvard-Smithsonian Center for Astrophysics, 60 Garden Street, 
Cambridge, MA 02138}
}









\begin{abstract}

We present stellar mass measurements for a sample of 36 Luminous
Compact Blue Galaxies (LCBGs) at redshifts $z$ = 0.4-1.2 in the
Flanking Fields around the Hubble Deep Field North. The technique is
based on fitting a two-component galaxy population model to
multi-broadband photometry. Best-fit models are found to be largely
independent on the assumed values for the IMF and the metallicity of
the stellar populations, but are sensitive to the amount of extinction
and the extinction law adopted. On average, the best-fit model
corresponds to a LMC extinction law with E(B-V)=0.5.  Stellar mass
estimates, however, are remarkably independent on the final model
choice.
Using a Salpeter IMF, the derived median stellar mass for this sample
is $5\times10^9$ M$_\odot$, i.e., $\sim$2 times smaller than
previous virial mass estimates.
Despite uncertainties of a factor 2-3, our results strengthen prior 
claims that L$^\ast$ CBGs at intermediate redshifts are, on average, 
about 10 times less massive than a typical L$^\ast$ galaxy today.

\end{abstract}




\section{Introduction}\label{intro_sec}

Luminous Compact Blue Galaxies (LCBGs) refers to a heterogenous class
of starburst systems with typical luminosities around $L^\star$ that
become very common at intermediate redshifts (Koo et al. 1994, 1995;
Guzm\'an et al. 1996, 1997; Phillips et al. 1997; Mall\'en-Ornelas et
al. 1999; Hammer et al. 2001). LCBGs are particularly interesting for
studies of galaxy formation and evolution since they have evolved more
than any other galaxy class in the last $\sim$8 Gyrs (Mall\'en-Ornelas
et al. 1999), and have been identified as a major contributor to the
observed enhancement of the global star formation rate density of the
universe at $z\le1$ (Guzm\'an et al. 1997; Gruel 2002).  Despite their
cosmological importance, the nature of LCBGs and their relation to
today's galaxy population still remain largely unknown.  The most
comprehensive study of LCBGs at intermediate redshift to date --only
45 objects-- is that of Phillips et al. (1997) and Guzm\'an et al.
(1997), who concluded that the LCBG class is populated by a mixture of
starbursts. About $\sim$60\% of galaxies in their sample are
classified as ``HII-like'' since they are spectroscopically similar to
today's population of vigorously star-forming HII galaxies. The
remaining $\sim$40\% are classified as ``SB disk-like'' since they
are more evolved star-forming systems similar to local spirals
with a central starburst, and giant irregular galaxies.

Given the diverse nature of the LCBG population, they are unlikely to 
evolve into one homogeneous galaxy class. 
There are two popular scenarios. Koo et al. (1994) and
Guzm\'an et al. (1996) have suggested that some subset of the most
compact HII-type LCBGs at intermediate redshifts may be the
progenitors of local low-mass spheroidal galaxies (or dwarf
ellipticals), such as NGC 205. 
Alternatively, Phillips et al. (1997) and Hammer et al. (2001) have
suggested that SB disk-like LCBGs may actually be more massive disks 
forming from the center outward to become present-day $L^\star$ spirals.
At the heart of this debate is the question: are the masses of LCBGs
comparable to that of today's massive or low-mass galaxies?  Prior
measures of Keck velocity widths and HST sizes show that LCBGs have
mass-to-light ratios that are only $\sim 0.1-1.0 M_\odot/L_{B\odot}$,
i.e., about 10 times smaller than a typical $L^\star$ galaxy today
(Koo et al. 1994; Guzm\'an et al. 1996; Phillips et al. 1997). Since
ionized gas emission line widths may not reflect the true
gravitational potential, an independent estimate of galaxy mass is
desirable. Stellar mass is an obvious choice, since HI and CO emission
line widths are too difficult to measure in galaxies beyond the local
universe. 

Several authors have pointed out that the near-IR luminosity of a
galaxy is a robust estimator of its stellar mass (e.g.,
Arag\'on-Salamanca et al. 1993; Alonso-Herrero et al. 1996; Charlot
1998). The main uncertainty in the inferred stellar masses arises from
the age of the stellar population (Rix \& Rieke 1993). Instead of
using H$\alpha$ to constrain the strength and age of a young burst
(Gil de Paz et al. 2000; \"Ostlin et al. 2001; P\'erez-Gonz\'alez et
al. 2003a, 2003b), an alternative approach when studying galaxies at
higher redshifts is to use rest-frame far-UV fluxes (Brinchmann \&
Ellis 2000, hereafter BE00; Papovich, Dickinson \& Ferguson 2001).
Here we use a modified version of the method described in BE00 to
estimate the stellar masses of a sample of LCBGs at intermediate
redshift.  Our goal is to test previous claims that LCBGs have, on
average, an order of magnitude smaller masses than typical galaxies
today with similar luminosities. In section 2 we introduce the
photometric data set used in this analysis. Section 3 describes the
technique developed to derive stellar masses. The results are
summarized in Section 4. For consistency with previous work,
throughout this paper we adopt Ho = 50 km s$^{-1}$ Mpc$^{-1}$ and qo =
0.05.

\section{The Data}\label{data_sec}

	The initial galaxy sample consists of 40 LCBGs at redshifts
$0.4<z<1.2$ selected from the Flanking Fields around the Hubble Deep
Field North, as described in Phillips et al. (1997). A wide variety of
information for this sample was made publicly available by the Deep
Evolutionary Extragalactic Probe (DEEP) collaboration, including F814W
surface photometry, (V$_{F606W}$ - F814W) colors, Oxygen and Balmer
emission line equivalent widths, velocity widths, star formation
rates, and virial masses (Phillips et al. 1997; Guzm\'an et
al. 1997). In addition, deep optical Un, G, R photometry for the
entire HDF+FF area is available through the Caltech Faint Galaxy
Redshift Survey (Hogg et al. 2000), while deep near-IR photometry in
the notched HK' filter is listed in the Hawaii Flanking Field Catalog
(Barger et al. 1999). After cross-correlating the various catalogs, we
ended up with a final sample of 36 objects with deep Un, G, R, F814W,
and HK' photometry. Two thirds of the objects are classified as
HII-like, while the remaining one third are classified as SB
disk-like.  Limiting magnitudes for the photometry in each of these
bands are: Un = 25, G = 26, R = 25.5, F814W = 25.0, HK' = 20.7. As
described in Phillips et al. (1997), all LCBGs in our sample are
brighter than F814W=23.74 mag and their half-light radii are smaller
than 0.5 arcsec. For F814W and HK', an estimate of the
total magnitude was derived from the total flux within a 3 arcsec
diameter aperture.  For Un, G, and R, total magnitudes were
derived after correcting the magnitudes within 1.7 arcsec diameter
apertures for flux outside the aperture considering the point source
case, as described in Hogg et al. (2000). The largest errors in the
combined data set are due to the comparatively shallower near-IR data
(typically $\sigma_{HK'}\sim0.2$ mag).


\section{Stellar Mass Estimates}\label{vp}

	The resulting photometric data set covers Un through HK' in
the observed frame. The red end of this range corresponds to
$\sim$0.9$\mu$m in the rest-frame of the highest redshift considered
here, and to $\sim$1.3$\mu$m for a more typical $z\sim0.7$ galaxy in
our sample.  The blue end corresponds to rest-frame $\sim$0.18$\mu$m
and $\sim$0.23$\mu$m, respectively. Gil de Paz \& Madore (2002) have
shown that this filter combination is well suited to derive physical
properties of galaxies at $z\sim0.7$ from broad-band data alone,
including star formation timescale, age, metallicity, and stellar
mass.

	Similarly to the stellar mass estimates method described in
BE00, we adopt a grid of evolutionary synthesis models with simple
star formation histories (Bruzual \& Charlot 2002) 
and select the most appropriate evolutionary history by
error-weighted fitting of the optical and IR photometry. The models
have three main parameters: initial mass function (IMF), star
formation rate (SFR) history, and metallicity. There are, however,
some differences between our method and that described by
BE00. Firstly, instead of k-correcting the observed magnitudes for our
LCBG sample, we chose to shift all model spectra to the redshift
of each of our objects.
Redshifted model spectra were convolved with the empirical filter
functions corresponding to the actual Un, G, R, F814W, and HK'
broadband filters used in the observations to derive the model
magnitudes. These magnitudes were then corrected for extinction
assuming not only a range of values for the color excess E(B-V) a la
BE00, but also a variety of extinction laws including those derived
for the Milky Way (MW; Mathis 1990), Large Magellanic Cloud (LMC;
Bouchet et al. 1985), LMC with wavelength-dependent extinction
correction (LMC-modified; Mas-Hesse \& Kunth, private communication),
Small Magellanic Cloud (SMC; Bouchet et al. 1985), and the so-called
Calzetti's law (C94; Calzetti et al. 1994).

	Secondly, there is growing evidence that LCBGs, both in the
local and distant universe, have a composite stellar population that
can be well represented by a young burst superimposed on an older
underlying population (Guzm\'an et al. 1998; Bergvall \& \"Ostlin
2001; \"Ostlin et al. 2001; Gruel 2002). We use a composite
``underlying+burst'' model to fit the observed broadband
magnitudes. The burst and underlying components are defined in terms
of the assigned star formation rate histories, e.g. instantaneous
burst versus constant or exponentially decaying SFR.  For simplicity,
both populations are assumed to have the same IMF and metallicity. The
best fit model is found using a two-step, iterative
procedure. Firstly, a single burst model is fitted to the observed Un,
G, and R optical bands alone, i.e., the extinction-corrected model
magnitudes are compared to the observed magnitudes and the best fit is
selected using a simple, weighted, least-squares minimization
technique.\footnote{No correction for nebular continuum were made to
the Un, G, and R magnitudes since its contribution is only significant
for burst ages of a few $10^6$ yr which are ruled out by the H$\beta$
equivalent widths measured for our sample galaxies (Guzm\'an et
al. 1997). The correction for nebular line emission on these broadband
magnitudes is estimated to be $<10$\% (Bergvall \& \"Ostlin 2002) and
were not applied.} Weights were scaled with the flux in each band
using the limiting magnitudes to provide the reference minimum
weight. Since the observed Un, G, and R bands map approximately the
rest-frame UV region of the spectrum at the redshifts of our galaxies,
light is entirely dominated by the youngest stars in the burst. The
fitted model will provide an initial value for the age of the
burst. Secondly, we fit a ``underlying+burst'' model to all observed
magnitudes, including F814W and HK'. The burst component is now
restricted to vary by less than 10 timesteps (or a factor $\sim$2) in
age from the initial value.  Stellar masses are estimated by
normalizing the combined ``underlying+burst'' spectral energy
distribution to the observed magnitudes modulo the IMF. Table 1
summarizes the model ingredients.

	Over 800 different model combinations were investigated.  As
shown by Gil de Paz \& Madore (2002), the comparison of broad-band
photometry with the predictions of evolutionary synthesis models to
derive galaxy physical properties yields, in general, highly
degenerate results. A
first set of ``best fit'' models were selected on the basis of having
the lowest residuals, i.e., the lowest standard deviations for the
five differences between model and observed
magnitudes.\footnote{During the model fitting procedure, it was found
that offsets needed to be applied to the observed magnitudes. 
We have not been able to identify the source of these offsets. They
were derived iteratively as part of the fitting technique. Extensive
Montecarlo simulations were performed to ensure that our code could
identify and recover artificially introduced zero-point offsets in
fake galaxy catalogs that best mimick the properties of our LCBG
sample. We are confident that the uncertainty in the derived stellar
masses is dominated by model degeneracy and random photometric errors,
as explained in the text, and not by offset errors, which contribute
only 0.1-0.15 dex to the mass estimates.}
The median residuals of the ``best fit'' models are less than 0.15
mag. The largest median residuals are $\sim$0.35 mag, which correspond
to models that use the MW extinction law with E(B-V)=1. On this basis
alone, we were able to reject the following cases: (i) models with
constant SFR for the underlying component; (ii) models with MW and
LMC-modified extinction laws; and (iii) models with LMC and SMC laws
and E(B-V)$>$0.5. The residuals were found to be only slightly
dependent on the choice of IMF, metallicity, and star formation
timescales. Since there is growing evidence that the IMF in
star-forming and starburst regions is universal and consistent with
Salpeter's (Leitherer 1998), we adopt a Salpeter IMF with lower and
upper mass cutoffs of 0.08 and 125 M$_\odot$, respectively, for both
the burst and underlying components.
We also adopt a metallicity $Z=0.4Z_\odot$, consistent 
with the range of metallicities observed in distant LCBGs
(Guzm\'an et al. 1996; Kobulnicki \& Zaritski 1999). Finally, we set the 
star formation timescales to be instantaneous for the burst, and 
exponential with $\tau=1$ Gyr for the underlying component.

	The more restrictive choice of parameters reduces
significantly the number of ``best fit'' models. The solutions are,
however, still degenerate since various combinations of E(B-V), ranging
from 0 to 0.5 mag, and extinction laws (LMC, SMC, and C94), all produce
a low residuals fit to the data. 
Since the timescales for the burst and underlying components in the
models are now fixed, we find that the dominant source of degeneracy
is the age-extinction degeneracy (Gil de Paz \& Madore 2002). This is
in the sense that an older burst with less dust-extinction may yield
as good fit to the data as a younger but more extincted burst.  If the
burst is assumed to be less than $10^8$ yrs old, as suggested by the
relatively strong [OII]$\lambda$3727 and H$\beta$ equivalent widths
(Guzm\'an et al. 1997), then we find only one ``best-fit'' model that
fulfills all the conditions described in our analysis. According to
this model, LCBGs are consistent with being experiencing, on average,
a 9\% burst of star formation that is only 13 Myrs old. This
young component is superimposed on an older, redder population that
is, on average, 2 Gyrs old. These results are qualitatively in good
agreement with the burst strength and age estimates based on
mass-to-light ratios and color analysis for a different sample of
LCBGs (Guzm\'an et al. 1998).  The derived amount of extinction,
E(B-V)=0.5 mag, is also in good agreement with the average value
measured in starburst galaxies similar to LCBGs (Gil de Paz et
al. 2000; Hammer et al. 2001; Rosa-Gonz\'alez et al. 2002).  We
emphasise, however, that the results on the stellar mass measurements
are remarkably independent on the final model choice, as it is
discussed below.

	Figure 1 shows the distribution of stellar masses for our LCBG
sample.  The median value of the overall distribution is $5\times10^9$
M$_\odot$. This value is very similar to that derived for nearby LCBGs
(\"Ostlin et al. 2001). Note, however, that this value is about $\sim10$ 
times smaller than the average stellar mass in Hammer et al.'s (2001) 
sample of compact galaxies at intermediate redshifts. This difference is
most likely due to the intrinsic higher luminosity of Hammer et al.'s
sample, and the fact that these authors adopted a different K-band 
mass-to-light ratio. The median values for the HII-like and SB
disk-like LCBGs are $5\times10^9$ M$_\odot$ and $2\times10^{10}$
M$_\odot$, respectively. Although non-significant in this small
sample, these values suggest that SB disk-like LCBGs tend to be more
massive than HII-like LCBGs.  The mean value for the subsample of 17
LCBGs with blue luminosities around L$^*$ (i.e., $M_B^* \pm 1$) is
$1.8\pm1 \times10^{10}$ M$_\odot$. This value is one order of
magnitude lower than the stellar mass expected for a local $L^*$
galaxy. Assuming $M_K^* = -25.1$ (for Ho= 50 km s$^{-1}$ Mpc$^{-1}$;
Mobasher, Sharples \& Ellis 1993), and a K-band mass-to-light ratio of
1 M$_\odot$/L$_{K,\odot}$ (H\'eraudeau \& Simien 1997), the estimated
stellar mass is $2\times10^{11}$ M$_\odot$.  Thus, we conclude that
the L$^*$ CBGs at intermediate redshifts are about 10 times less
massive than a typical L$^*$ galaxy today. This result is in good
agreement with previous claims based on virial mass estimates (Koo et
al. 1994; Guzm\'an et al. 1996; Phillips et al. 1997).  Our conclusion
is not affected by the final choice of the model we adopted here. The
median value for the stellar mass estimates derived from the other
``best-fit'' models differ in a factor $\sim$2 from the median value
given above. The highest values correspond to models using the C94
extinction law, which yield a median stellar mass of
$1.3\times10^{10}$ M$_\odot$. We also note that a Salpeter IMF yields
systematically higher masses, although only by $0.2\pm0.05$ dex. The
use of a different IMF would strengthen our conclusion.  Most
interestingly, we find that LCBGs at intermediate redshifts span the
same range of stellar masses and stellar mass-to-light ratios as that
characteristic of Lyman-break galaxies at redshifts $z\sim3$
(Papovich, Dickinson \& Ferguson 2001).

	Figure 2 shows the comparison of the stellar masses with
virial masses derived by Phillips et al (1997). The virial masses used
for this figure have been corrected for a systematic underestimate of
the galaxy gravitational potential when emission-line velocity widths
are used.  This effect amounts to a $\sim$30\% increase in observed
velocity widths (Guzm\'an et al 1996; Rix et al 1997; Phillips et al
1997; Pisano et al 2001). No corrections for inclination were made.
Rotationally-supported, face-on LCBGs are likely to have virial masses
significantly lower than their stellar masses. The 1-$\sigma$ error
bars have been calculated by propagating the magnitude errors, the
standard deviations of the fit residuals, and the quoted errors in
half-light radii and velocity widths. The largest source of
uncertainty is associated to the errors in the shallower HK'
photometry. The average error in the stellar mass measurements amounts
to 0.36 dex (in logM). Excluding the two most uncertain measurements,
the average of the estimated errors drop to 0.27 dex, i.e., a factor
of $\sim$2. This value is comparable to the median error estimated by
BE00 in their error analysis. The average random error for the virial
mass measurements is estimated to be 0.18 dex. The median
virial-to-stellar mass ratio is 0.39 dex. Thus the stellar mass
estimates of LCBGs are, on average, $\sim$2 times smaller than the
virial mass estimates. This result is also in good agreement with the
comparison between stellar masses and dynamical masses made by BE00
and Papovich, Dickinson \& Ferguson (2001). The scatter in the
virial-to-stellar mass ratio is slightly larger than expected from the
error analysis, although the exclusion of the two most deviant points
results in 0.54 dex, or a scatter of a factor $\sim$3. No systematic
differences in the virial-to-stellar mass ratios are detected between
the SB disk-like and HII-like LCBGs. We note, however, that Figure 2
suggests a trend in the sense that the lowest stellar mass LCBGs tend
to have the highest virial-to-stellar mass ratios. 

\section{Conclusions}\label{conc_sec}
 
	We have presented a new method for determining robust stellar
masses for LCBGs up to redshifts $z\sim1.2$. Our method expands on
previous work in the field by including a two-component
``burst+underlying'' model population fitted to a combination of
rest-frame UV, optical and near-IR broad-band photometry in an
iterative fashion. Although specifically tailored for starburst
galaxies with a composite stellar population, this method can also be
used to estimate the stellar masses of the field galaxies in general
of known redshift.  According to our best-fit model, LCBGs at
intermediate redshifts can be described, on average, as starburst
systems which are undergoing a 13 Myrs old burst involving 9\% of the
galaxy mass superimposed on an underlying, older population (2
Gyrs). This model requires a modest amount of extinction (E(B-V)=0.5)
with a LMC extinction law. We have demonstrated via comparisons and
simulations that the stellar masses can be estimated with a precision
of log $\Delta$M$_{star}$ $\sim$0.3 dex, and are not significantly
affected by the main assumptions needed to constrain the degeneracy
intrinsic to this method. The median stellar mass for LCBGs in our
sample is $5\times10^9$ M$_\odot$, or $\sim$2.5 times smaller than
previous mass estimates based on the virial theorem using scale
lenghts and emission-line velocity widths. The new, independent mass
estimates are consistent with previous claims that L$^\ast$CBGs at
intermediate redshifts are about 10 times less massive than a typical
L$^\ast$ galaxy today.


\acknowledgments

RG is grateful to the Institut d'Astrophysique de Paris and the
Laboratoire d'Astrophysique at the Observatoire Midi-Pyrenee for their
hospitality and financial support for this project. RG also
acknowledges funding from NASA grants HF01092.01-97A and LTSA
NAG5-11635. G.O. acknowledges support from the Swedish Natural
Sciences Research Council and the STINT foundation. We thank Roser
Pell\'o for providing k-corrections used for an independent test of
the quality of the photometric data, and Jean Michel Desert for
helping in the crosscorrelation of the various catalogs.

\newpage

\section{References}

\begin{itemize}


\item[] Alonso-Herrero A., Arag\'on-Salamanca A., Zamorano J. \& 
Rego M. 1996, MNRAS, 278, 417

\item[] Arag\'on-Salamanca A., Ellis R.S., Couch W.J. \& Carter D. 
1993, MNRAS, 262, 794

\item[] Barger A.J., Cowie L.L., Trentham N., Fulton E., Hu E.M., 
Songaila A. \& Hall D. 1999, AJ, 117, 102

\item[] Bergvall N. \& \"Ostlin G. 2002, A\&A 390, 891

\item[] Bouchet et al. 1985, A\&A, 149, 330

\item[] Brinchmann J. \& Ellis R.S. 2000, ApJL, {536}, {77}

\item[] Bruzual G. \& Charlot S. 2002, in prep.

\item[] Calzetti et al., 1994, ApJ 429, 582

\item[] Charlot S. 1998, in Benvenuti P. et al, eds., AIP Conf. 
Proceedings Vol. 408, The Ultraviolet Universe at Low and High 
Redshift: Probing the Progress of Galaxy Evolution. American 
Institute of Physics, New York, p. 403

\item[] Gil de Paz A., Arag\'on-Salamanca A., Gallego J.,
Alonso-Herrero A., Zamorano J. \& Kauffmann G. 2000, MNRAS, 316, 357

\item[] Gil de Paz A. \& Madore B.F. 2002, AJ, 123, 1864

\item[] Gruel N. 2002, Ph.D. thesis, Observatoire de Paris

\item[] Guzm\'an, R., Koo, D. C., Faber, S. M., Illingworth, G. D.,
Takamiya, M., Kron, R., \& Bershady, M. A. 1996, ApJ, {460}, {L5}

\item[] Guzm\'an R., Gallego J., Koo D. C., Phillips A. C.,
Lowenthal J. D., Faber S. M., Illingworth G. D. \& Vogt N. P. 1997,
ApJ, {489}, {559}

\item[] Guzm\'an R., Koo D. C., Jangren, A., Bershady M., Faber
S. M. \& Illingworth G. D., 1998, ApJ, {495}, {L13}

\item[] Hammer, F., Gruel, N., Thuan, T. X., Flores, H., \&
        Infante, L., 2001, ApJ, {550}, {570}

\item[] H\'eraudeau P. \& Simien F. 1997, A\&A, 326, 897

\item[] Hogg, D.W., Pahre M.A., Adelberger K.L., Blandford R., Cohen J.G.,
Gautier T.N., Jarrett T., Neugebauer G. \& Steidel C.S. 2000, ApJS, 127, 1

\item[] Kobulnicky H. D., \& Zaritsky D. 1999, ApJ, {511}, {118}

\item[] Koo, D. C., Bershady, M. A., Wirth, G. D., Stanford, S. A., \&
        Majewski, S. R. 1994, ApJL, {427}, {L9}

\item[] Koo, D. C. Guzm\'an, R., Faber, S. M., Illingworth, G. D.,
Bershady, M. A., Kron, R., \& Takamiya, M. 1995, ApJ, 440, L49

\item[] Leitherer 1998, in Gilmore G., \& Howell D. eds., 38th
Herstmonceux Conference, The Stellar Initial Mass Function. ASP
Conference Series, Vol. 142, p.61

\item[] Leitherer et al. 1999, ApJS, 123, 3

\item[] Mall\'en-Ornelas, G., Lilly, S. J., Crampton D., \& Schade D.
1999, ApJ, {518}, {83}

\item[] Mathis 1990, ARA\&A, 38

\item[] Mobasher B., Sharples R.M. \& Ellis R.S. 1993, MNRAS, 263, 560

\item[] \"Ostlin G., Amram P., Bergvall N., Masegosa J., Boulesteix J.
\& M\'arquez I. 2001, A\&A 374, 800

\item[] Papovich C., Dickinson M. \& Ferguson, H.P. 2001, ApJ, 559, 620

\item[] P\'erez-Gonz\'alez P.G., Gil de Paz A., Zamorano J., Gallego J., 
Alonso-Herrero A. \& Arag\'on-Salamanca A. 2003a, MNRAS, 338, 508

\item[] P\'erez-Gonz\'alez P.G., Gil de Paz A., Zamorano J., Gallego J., 
Alonso-Herrero A. \& Arag\'on-Salamanca A. 2003b, MNRAS, 338, 525

\item[] Phillips A. C., Guzm\'an R., Gallego J., Koo D. C.,
Lowenthal J. D., Vogt N. P., Faber S. M. \& Illingworth G. D. 1997,
ApJ, {489}, {543}

\item[] Pisano D.J., Kobulnicky H. D., Guzm\'an R., Gallego J. \& 
Bershady, M. A. 2001, AJ, 122, 1194

\item[] Rix H.W. \& Rieke M.J. 1993, ApJ, 418, 123

\item[] Rix H.W., Guhathakurta P., Colles M. \& Ing K. 1997, MNRAS, 285, 779

\item[] Rosa-Gonz\'alez D., Terlevich E. \& Terlevich R. 2002, MNRAS, 
332, 283

\end{itemize}

\clearpage

\begin{deluxetable}{ll}
\tabletypesize{\scriptsize}
\tablecaption{Grid of Model Spectral Energy Distributions} 
\tablewidth{0pt}
\tablehead{
\colhead{Parameter} &
\colhead{Range} \\
}
\startdata
Age.................... & 10$^6$ to $2\times10^{10}$ yr in steps of 0.1 in log yr \\
SFR (burst) ...... & Instantaneous and $\tau$=1.0 Gyr \\
SFR (underlying)  & $\tau$=1.0 Gyr and constant\\
IMF.................... & Salpeter, Kennicutt\\
Metallicity ......... & 0.4 $Z_\odot$, and $Z_\odot$\\
E(B-V)............... & 0.0, 0.25, 0.5, 1.0 mag\\
Ext. Law ........... & MW, LMC, LMC-modified, SMC, C94\\
\enddata
\end{deluxetable}


\clearpage

\begin{figure}
\epsscale{0.75}
\plotone{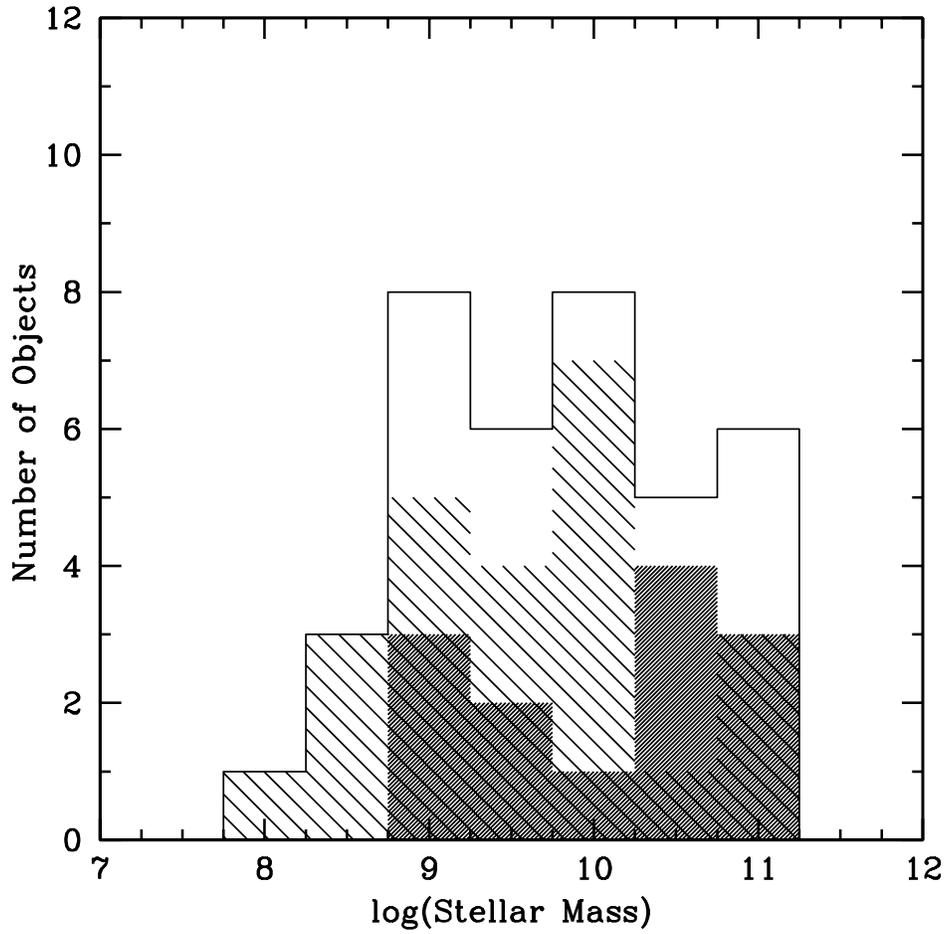}
\vskip 1 cm
\caption{Histogram of stellar masses for our sample of LCBGs at
intermediate redshift. Solid line: distribution of stellar masses for
the whole sample. Grey histogram: distribution of stellar masses for
SB disk-like LCBGs. Shaded histogram: distribution of stellar masses
for HII-like LCBGs.}

\end{figure}

\begin{figure}
\epsscale{0.75}
\plotone{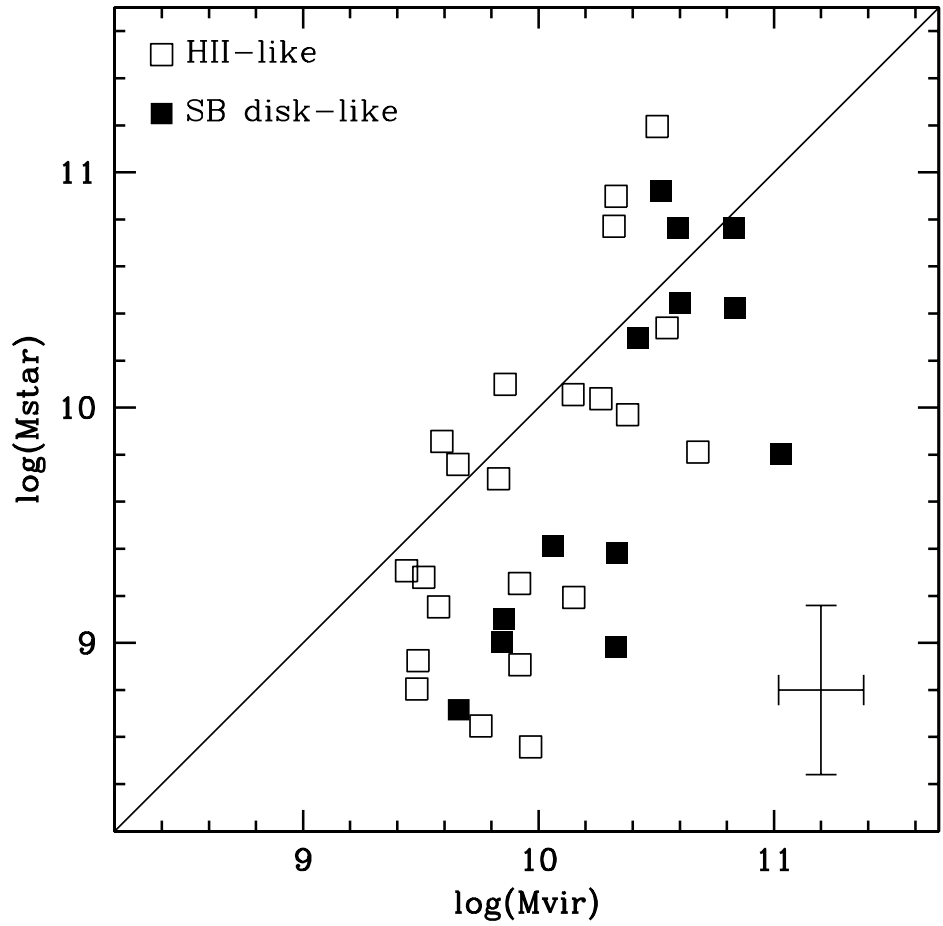}
\vskip 1 cm
\caption{The virial mass versus stellar mass for LCBGs in our sample 
in log(M$_\odot$) units. 
Open squares: HII-like LCBGs; solid squares: SB disk-like LCBGs.
Error bars correspond to average 1-$\sigma$ errors.}

\end{figure}

\end{document}